%% file: charm07_BrianMeadows.tex
\documentclass[twocolumn,twoside,slac_two,preprintnumbers]{revtex4}
\usepackage{graphicx}
\usepackage{fancyhdr}
\usepackage{epsfig}
\begin{document}

\title{\boldmath Low Mass $S$-wave $K\pi$ and $\pi\pi$ Systems}


\author{B. Meadows}
\affiliation{Department of Physics, University of Cincinnati, Cincinnati, OH 45221, USA}

\begin{abstract}
Knowledge of the details of the $S$-wave $K\pi$ and $\pi\pi$ systems 
limits the precision of measurements of heavy quark meson properties.  
This talk covers recent experimental developments in parametrizing
and measuring these waves, and examines possibilities for the future.
\end{abstract}

\preprint{UCHEP-07-09}
\maketitle

\thispagestyle{fancy}


\input{babar}
\input{mytex}
\newcommand{\mco}{\multicolumn}
\newcommand{\babar}{B{\small A}B{\small AR}}
\newcommand{\belle}{Belle}

\section{Introduction}
Hadrons were once regarded as a source of
complication in the measurement of the dynamics of weak decays of
heavy quark states.  The realization that their
interactions can be used to help in understanding the phases
involved is quite a recent development.  There are several examples.
Interference between hadrons produced in decays of the $\Dz$
and $\Dzbar$ mesons to the self-conjugate final state $\KS\pip\pim$
\footnote{Charge conjugate states are implied unless specified otherwise
throughout this talk}
has been used to measure $\Dz$-$\Dzbar$ mixing parameters.
\cite{Abe:2007rd,Asner:2005sz}
It has also allowed the measurement of the CKM phase $\gamma$ 
from the decays $B^+\to\Dz\Kp$ and  $B^+\to\Dzbar\Kp$.
\cite{Giri:2003ty,Aubert:2005iz,Poluektov:2006ia,Aubert:2007ii}
Interference between $S$- and $P$-waves in the $\Kp\pim$ systems
produced in $\Bz\to J/\psi\Kp\pim$ decays has been used to resolve
the ambiguity in the sign of $\cos 2\beta$, where $\beta$ is the 
CKM phase involved in these decays.

Central to all these analyses is the need to assign a partial
wave composition at each point in the hadron phase space.
Currently, models for the hadron interactions based on resonant
composition are mostly used to do this (the ``isobar model").
Sometimes a $K$-matrix description of the $S$-wave component
is included.  Uncertainties and ambiguities or questionable 
assumptions, especially with respect to the $S$-wave behaviour
lead to significant systematic uncertainties in the results.  
As the precision of the measurements improves, it becomes a more
pressing goal either to develop a good, model-independent
strategy for describing the hadron amplitudes or, at least, a
better understanding of how to describe the $S$-waves in the
decays of heavy quark mesons.

It is also of intrinsic interest to study the low mass
$S$-wave $K\pi$ and $\pi\pi$ systems to better understand the
scalar states, labelled here as $\kappa(800)$ or $\sigma(500)$,
that may exist
\cite{vanBeveren:2006ua}.
Measurements of these amplitudes in this region are sparse.
There is also merit, therefore, in pursuing the possibility of
using heavy quark meson decays to add experimental data in
these regions.  This talk focusses on recent progress and
future prospects on both fronts.  

In the first section, information
presently available on both $K\pi$ and $\pi\pi$ $S$-waves is
briefly reviewed.  The expected relationship between scattering
amplitudes and those measured in decays of heavy quark mesons is 
then examined.  The remainder of the talk discusses ways in 
which experimental observations, mostly of the available $K\pi$ 
data, are being studied.

\section{\boldmath $S$-wave Scattering Data}
Excellent experimental information comes from model-independent
analyses of differential cross sections for reactions in which 
production of $K\pi$ or $\pi\pi$ systems is dominated by a pion 
exchange mechanism.  The amplitudes so determined show clear 
evidence for the $K^*_0(1430)$ resonance in the iso-spin $I=1/2$ 
$K\pi$  $S$-wave and for the $f_0(980)$ in the $I=0$ $\pi\pi$ 
$S$-wave.  No exotic states are found in the $I=3/2$ $K\pi$ or
$I=2$ $\pi\pi$ waves.  Data at the very low mass regions, where 
$\kappa(800)$ or $\sigma(500)$ poles may lurk far from the real 
axis, are relatively poor.

\subsection{\boldmath The $K\pi$ System}

\begin{figure*}[hbt]
\centering
\epsfig{file=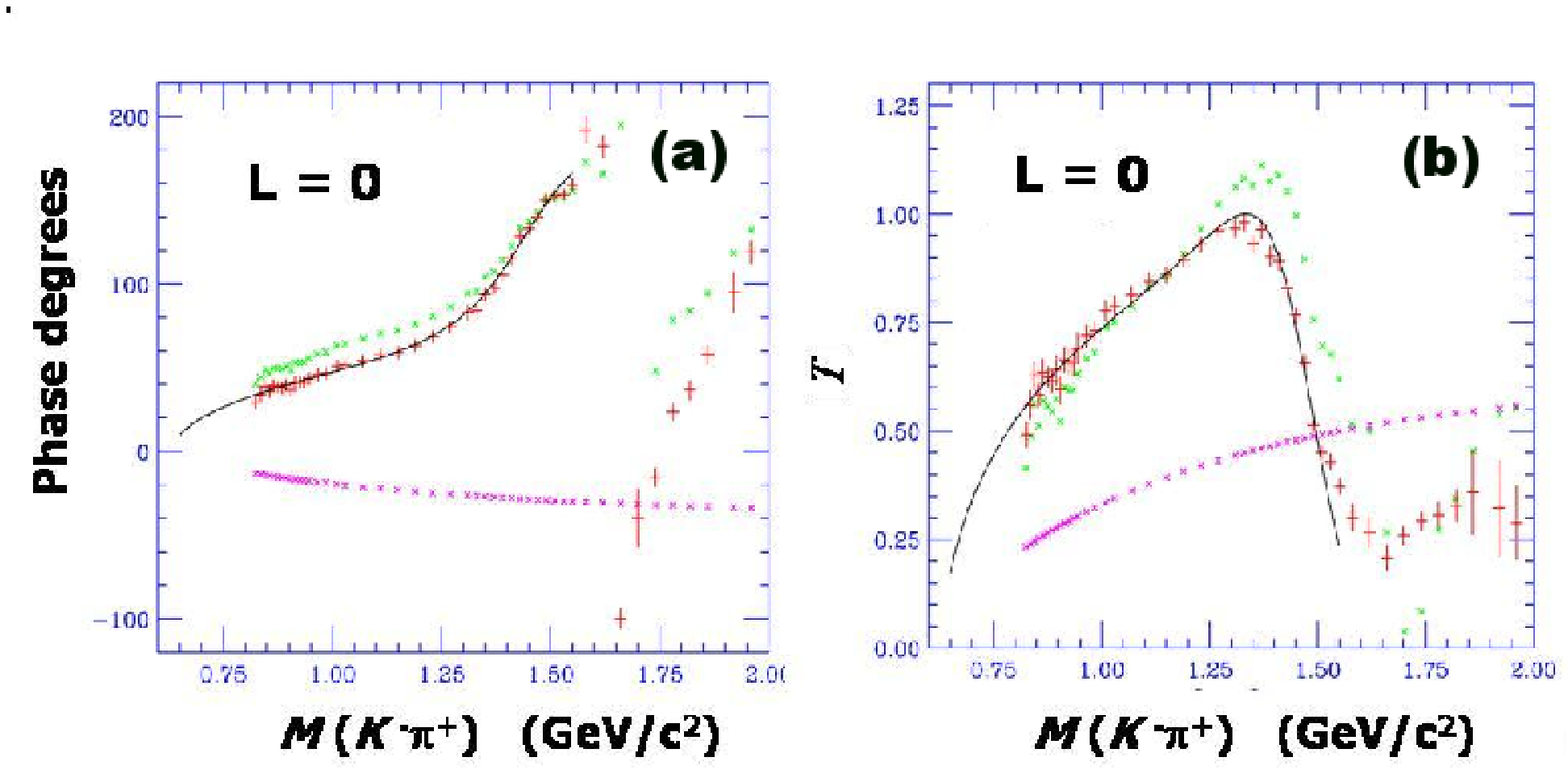,
          width=0.66\textwidth,
          angle=0}
\epsfig{file=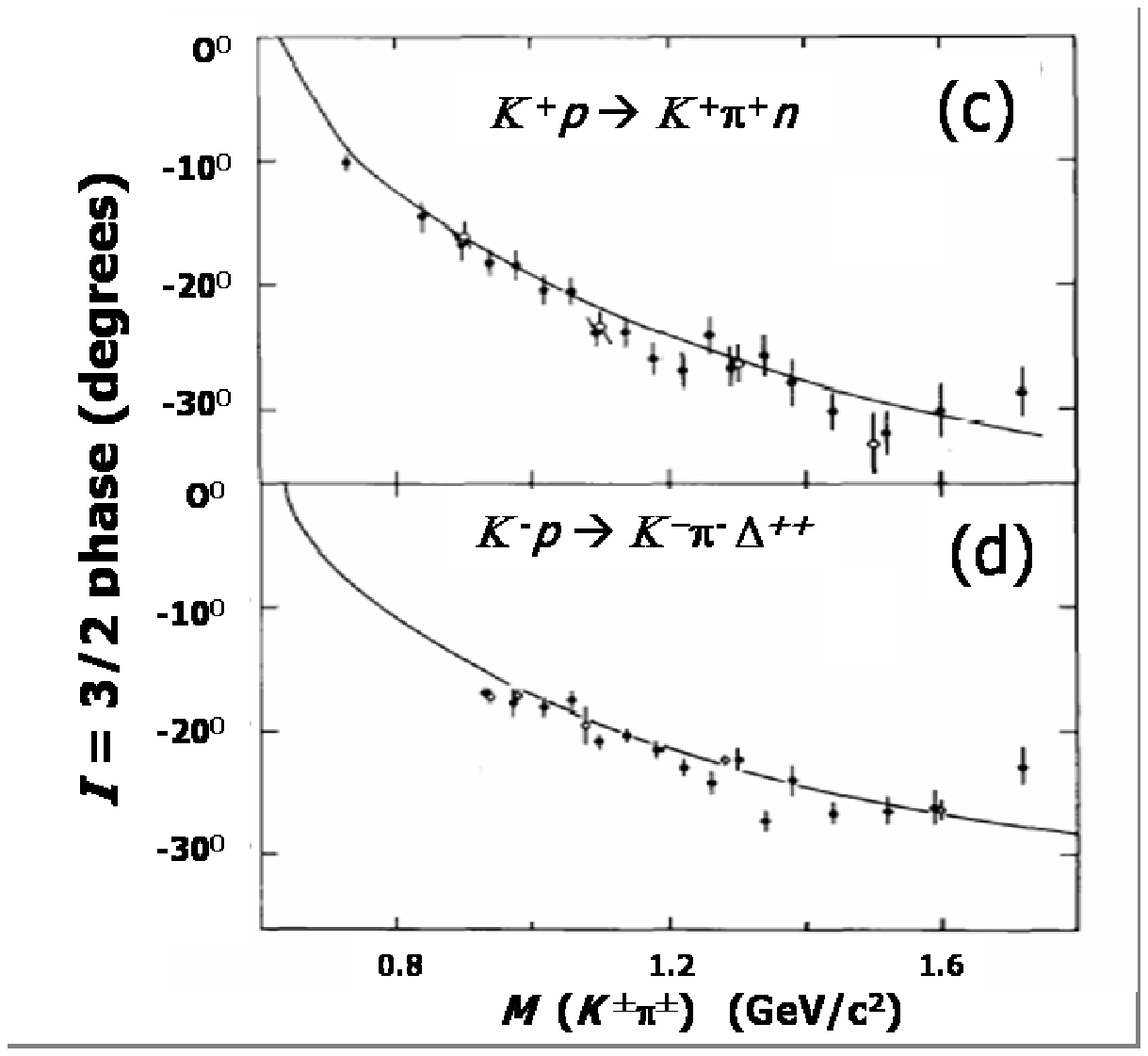,
          width=0.33\textwidth,
          angle=0}
\caption{$S$-wave $\Km\pip$ scattering amplitude extracted by the LASS
 collaboration from their data on the reaction $\Km p\to\Km\pip n$
 at 11~GeV/c.  The phase is shown in (a) and the magnitude in (b).
 The $I$-spin 1/2 component is represented by points on a solid curve
 and $I$-spin 3/2 is shown with no curve.
 $I=3/2$ phases measured in data from Ref.~\cite{Estabrooks:1977xe}
 are plotted as a function of $K^{\pm}\pi^{\pm}$ invariant mass ofr
 (c) $\Kp p\to\Kp\pip n$ and (d) $\Km p\to\Km\pim\Delta^{++}$.
\label{fig:LASS}
} 
\end{figure*}

The SLAC E135 (LASS) experiment
\cite{Aston:1987ir,wmd:private}
provides the best information on the $\Km\pip$ system.  The data 
are shown in Fig.~\ref{fig:LASS}.  Note that there are no data
below  825~MeV/c$^2$ from this experiment, though some is available
from an earlier era of experiments.  The LASS collaboration
determined that both $I=1/2$ and $I=3/2$ amplitudes $T(s)$ are
unitary up to $K\eta$' threshold ($\sqrt s=1454$~MeV/c$^2$)
\begin{eqnarray}
  T(s) = \sin\delta(s) e^{i\delta(s)}
  \label{eq:unitary}
\end{eqnarray}
where $s$ is the squared invariant mass and $\delta(s)$ is the 
phase.

The $I=3/2$ phase, assumed to be unitary, was measured by
Estabrooks, \etal\
\cite{Estabrooks:1977xe}
and is shown in Fig.~\ref{fig:LASS}(c) and (d).  The LASS
collaboration used this information to separate the $I=1/2$ and
$I=3/2$ components of the amplitude seen in
Fig.~\ref{fig:LASS}(a) and (b).  The fit to their data, shown
as the solid curves, are described by Eq.~(\ref{eq:unitary}) with 
\begin{eqnarray}
  \begin{array}{rrcl}
    {\boldmath I=1/2:}
           & \delta^{1/2}(s) &=& \delta_R^{1/2}(s)+\delta_B^{1/2}(s) \\
           & \cot\delta_R^{1/2}(s) &=& (s_0-s)/(\sqrt s_0\Gamma_0) \\
           & q\cot\delta^{1/2}_B(s) &=& 1/a_{1/2} + b_{1/2}q^2 \\[8pt]
    {\boldmath I=3/2:}
           & q\cot\delta^{3/2}(s) &=& 1/a_{3/2} + b_{3/2}q^2 \\
  \end{array}
 \label{eq:lasswave}
\end{eqnarray}
where $q$ is the momentum of the $\Km$ in the $K\pi$ center of mass.
This parametrization is valid in the range $825<\sqrt s<1454$~MeV/c$^2$,
and includes one $I=1/2$ resonance of mass
$\sqrt s_0\sim 1435$~MeV/c$^2$ and width $\Gamma_0\sim 275$~MeV/c$^2$.  
Non-resonant backgrounds in both waves are described by scattering
lengths $a$ and effective ranges $b$.

\subsection{\boldmath The $\pi\pi$ System}
\begin{figure}[hbt]
\centering
\epsfig{file=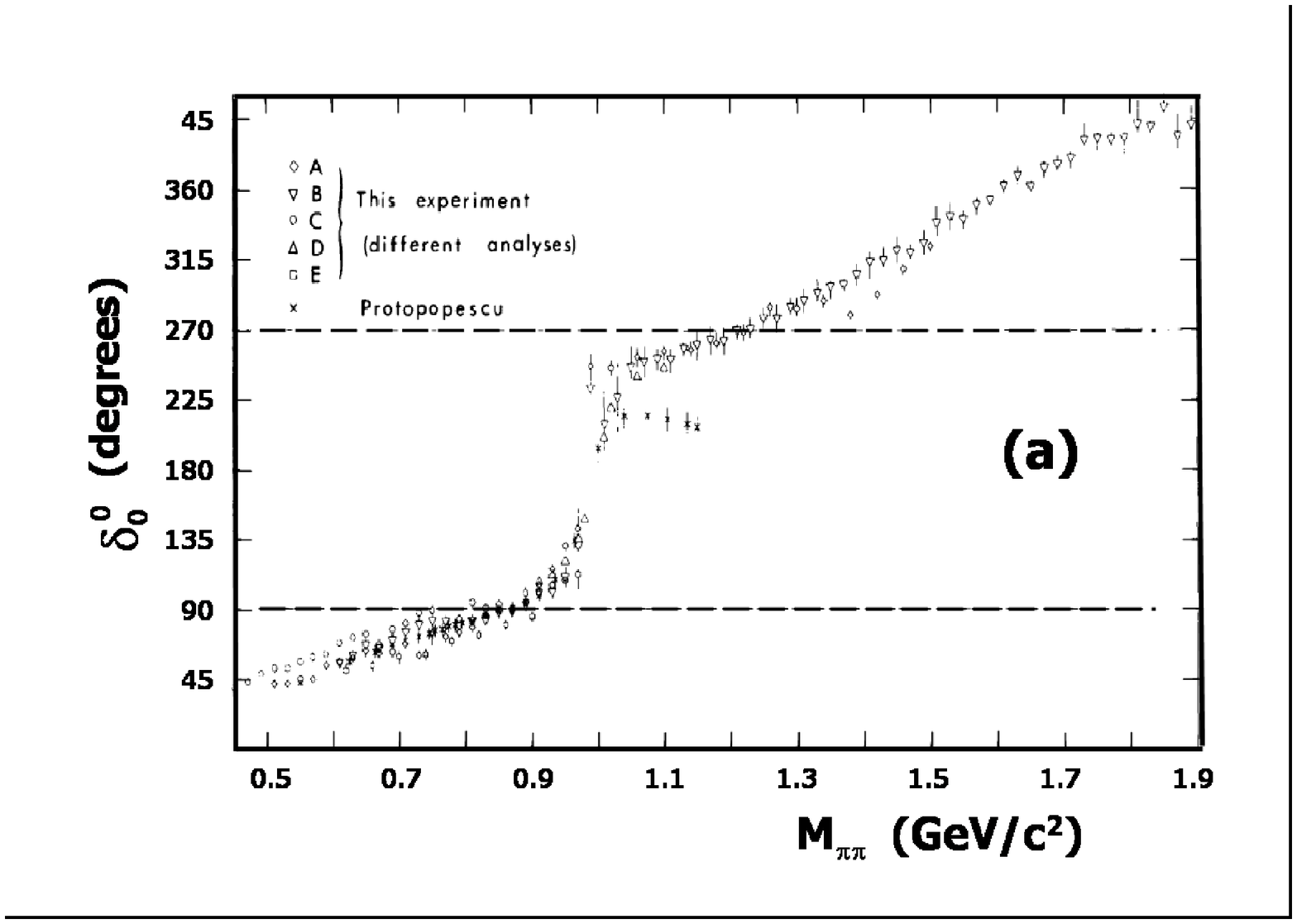,
          width=0.48\textwidth,
          angle=0}
\\
\epsfig{file=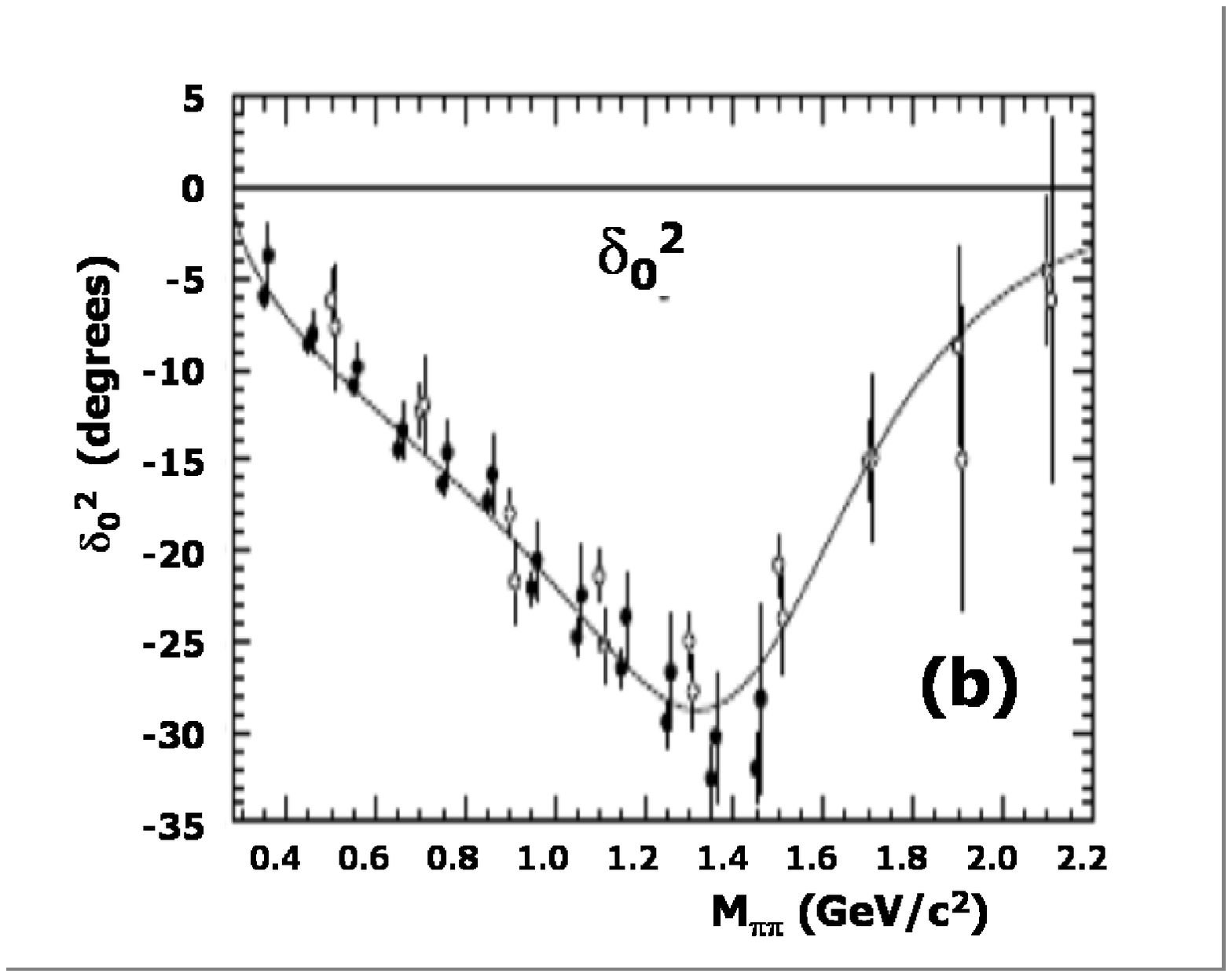,
          width=0.48\textwidth,
          height=0.25\textwidth,
          angle=0}
\caption{(a) $I=0$ phase of the $S$-wave $\pi\pi$ scattering amplitude
 extracted from data for the reactions $\pim p\to\pim\pip n$ from 
 Ref.~\cite{Grayer:1974cr}.  The phase in degrees is plotted as a
 function of
 $\pi\pi$ invariant mass $M$.  (b) The $I=2$ phase shifts plotted as
 a function of $\pi\pi$ invariant mass $M$.  The $I=2$ amplitude is
 assumed to be unitary up to 1500 MeV/c and is obtained from data for
 the reactions $\pip p\to\pip\pip n$ (12.5 GeV/c)
 \cite{Hoogland:1974cv}
 and $\pip d\to\pim\pim pp_{\hbox{spectator}}$ (9 GeV/c)
 \cite{Durusoy:1973aj}.
} \label{fig:Grayer}
\end{figure}

Phases for the $I=0$ component, shown in Fig.~\ref{fig:Grayer}(a), have
been extracted in several analyses of $\pim p\to\pim\pip n$ data from
Grayer, \etal
\cite{Grayer:1974cr}
It is clear that data below 600~MeV/c$^2$ in the region of the
$\sigma(500)$ from these data are poor.  

The $I=2$ amplitude, assumed to be unitary up to
$\rho\rho$ threshold ($\sim 1500$~MeV/c$^2$), was derived from data
on $\pip p\to\pip\pip n$ interactions at 12.5~GeV/c
\cite{Hoogland:1974cv}
and $\pip d\to\pim\pim pp_{\hbox{spectator}}$ at 9 GeV/c
\cite{Durusoy:1973aj}.
It was fit to the form shown in Fig.~\ref{fig:Grayer}(b)
\cite{Achasov:2003xn}.
For $I=0$, the amplitude is unitary up to $K\bar K$ threshold where its
elasticity drops suddenly.  Slightly below this, the phase rises rapidly
indicating the presence of the $f_0(980)$ resonance.

\section{\boldmath Role of Scattering in $D$ Decays}
In terms of our two goals, it would be useful if we could draw on
these measurements to help in reducing ambiguities in models used
in analyzing decays of heavy quark mesons.  At the same time, we
need to learn how to interpret such decays in learning more about
the scattering amplitudes at small $S$ values.

Consider the decay $D\to (AB)C$ in which a two hadron system $f=AB$
and another system $C$ are produced.  A simple assumption about the
decay amplitude $F(s)$ for such a process is that it can be
factorized
%%%, as illustated in Fig.~\ref{fig:Watson}, 
into short and
long-range effects.
\begin{equation}
  F_f(s) = T_{fk}(s)Q_k(s)
  \label{eq:watson}
\end{equation}
Here $Q_k(s)$ describes the short-range decay of $D$ to $C$ and an
intermediate hadron system $k$, and $T_{kf}(s)$ describes the
subsequent re-scattering within the system $k$ to produce the final
state $f$.  We take $Q_k(s)$ to encode not only the relatively real
ratio of decay modes for the weak decay of the $D$ that would imply
an $s$-dependence of its magnitude, but also any other short-range
effects that may occur in the $C k$ system which might also
impart an $s$-dependence of its phase.

%%\begin{figure}[hbt]
%%\centering
%%\includegraphics[width=0.35\textwidth]{Figures/figure3.eps}
%%\caption{A $\Dp\to\Km\pip\pip$ decay.  
%%} \label{fig:Watson}
%%\end{figure}

If $s$ is below the threshold where $AB$ scattering becomes
inelastic ($K\eta'$ for $L=$ even waves when $AB=K\pi$ and
$\Kp\Km$ when $AB=\pi\pi$ systems, for example), then $T_{kf}(s)$ is
simply equal to the $AB$ elastic scattering amplitude.  If the phase
of $Q_k(s)$ is independent of $s$, then the phase of $F(s)$
will have the same $s$-dependence as $T(s)$, \ie\ that observed in
elastic scattering.  This is a statement of the (more rigorously
derived) Watson theorem
\cite{Watson:1952ji}.
If partial wave expansions of $F(s)$ and $T(s)$ are made, then
this condition must hold for each partial wave.  It must hold,
in particular, for the $S$-waves.

The Watson theorem could provide a very useful constraint, therefore,
in the analysis of heavy quark states, so its range of applicability
is of great importance.  The considerations above lead us to expect
the following:
\begin{itemize}
 \item
 If $C$ is a di-lepton ($\ell\nu$, for example) then $Ck$ scattering
 would be unlikely and the phase of $Q_k$ would have little, if any,
 $s$-dependence.
 \item
 The same might also be true if $C$ were a massive hadron (such as a
 $J/\psi$) with a small interaction radius.
 \item 
 In cases where $C$ is a hadron with mass comparable to $A$ or $B$,
 significant scattering in the $C k$ system Would be likely.  This
 could lead to a dependence of both the magnitude and the phase
 of $Q_k(s)$ upon $s$.
 \item
 The likelihood of $Ck$ scattering probably increases at shorter 
 range.  
\end{itemize}

We should expect, therefore, to find any deviations from the Watson
theorem to be most pronounced when the mass of $C$ is comparable with
that of $A$ or $B$, and at small $s$ in the $AB$ system.  We also note
that, the Watson theorem applies to every partial wave, and it makes
little sense to impose it only on the $S$-wave alone.

\section{\boldmath $K\pi$ Production from Heavy Quark Meson Decays}
\subsection{Decays with Leptons or Massive Hadrons}
\begin{figure}[hbt]
\centering
\epsfig{file=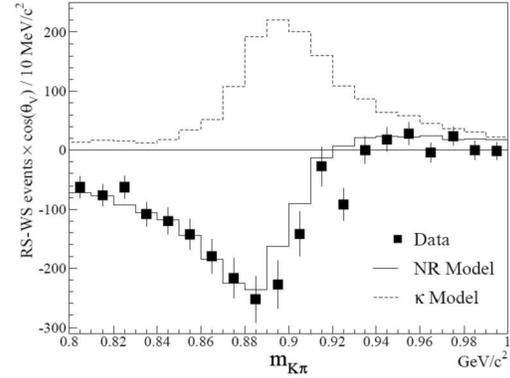,
          width=0.40\textwidth,
          angle=0}
\caption{Asymmetry in cosine of $\Km\pip$ helicity angle in
 $\Dz\to\Km\pip\ell\nu$ data from the FOCUS collaboration
 \cite{Link:2005ge}
 as a function of invariant mass for the $\Km\pip$ system.
 The behaviour observed in the LASS data is indicated by the
 solid curve.
} \label{fig:Focuslnu}
\end{figure}

In a study of semi-leptonic decays of $D$ mesons
\begin{eqnarray}
  \Dz &\to&\Km\pip\ell\nu.
  \label{eq:kpilnu}
\end{eqnarray}
the FOCUS collaboration
\cite{Link:2005ge}
observed asymmetry in the distribution of $\cos\theta$, where
$\theta$ is the angle between $\Km$ and the $\ell\nu$ system in 
the $\Km\pip$ rest frame.  The asymmetry results from 
interference between $S$- and $P$-waves, and is proportional to 
the cosine of the relative phase $\gamma$ between them.  As seen 
in Fig.~\ref{fig:Focuslnu} this follows the behaviour observed in 
the LASS data quite closely as the $\Km\pip$ invariant mass moves 
through the $K^*(890)$ region.  This conforms to the first of our 
expectations.

\begin{figure}[hbt]
\centering
\epsfig{file=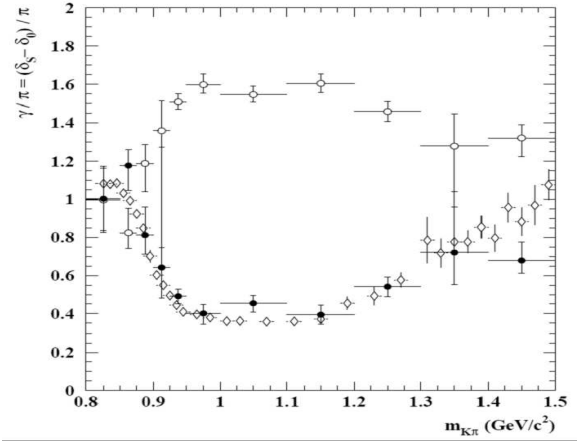,
          width=0.45\textwidth,
          height=0.35\textwidth,
          angle=0}
\caption{Two solutions (open and closed circles with error bars)
 for the relative phase $\gamma=\delta_S-\delta_P$ between $S$-
 and $P$-wave amplitudes of $\Kp\pim$ systems produced in
 $\Bz\to J/\psi\Kp\pip$ decays from the \babar\ collaboration
 \cite{Aubert:2004cp}.
 Values for $\gamma$ are plotted as a function of $\Kp\pim$
 invariant mass.
 The solid circles are for the only physical solution and follow
 the LASS data points, plotted as triangles, closely.
} \label{fig:BabarJpsikpi}
\end{figure}

The \babar\ collaboration observed 
\cite{Aubert:2004cp}
similar behaviour in the asymmetry of the helicity distribution 
for $\Kp\pim$ produced in
\begin{eqnarray}
  \Bz &\to& J/\psi\Kp\pim
  \label{eq:Jpsikpi}
\end{eqnarray}
decays.  Fig.~\ref{fig:BabarJpsikpi}
shows the two solutions for the $S$-$P$ phase difference $\gamma$
in the $K^*(890)$ invariant mass region.  The relative phase is
dominated by the rapid variation in the $P$-wave due to the
$K^*(890)$ and only one solution is physically meaningful.  It is
seen in the figure that this matches the LASS phase variation
(though shifted by $+\pi$ radians) very well.

This conforms to the second of our expectations, but it is not 
obvious why there is an overall phase shift of $+\pi$ radians.

\subsection{\boldmath Decays to Light Hadrons}
\label{sec:qphase}
The most detailed experimental information comes from studies of
$\Dp\to\Km\pip\pip$ decays.  These Cabibbo favoured decays are known
to contain a large $S$-wave component.

\begin{figure}[hbt]
\centering
\epsfig{file=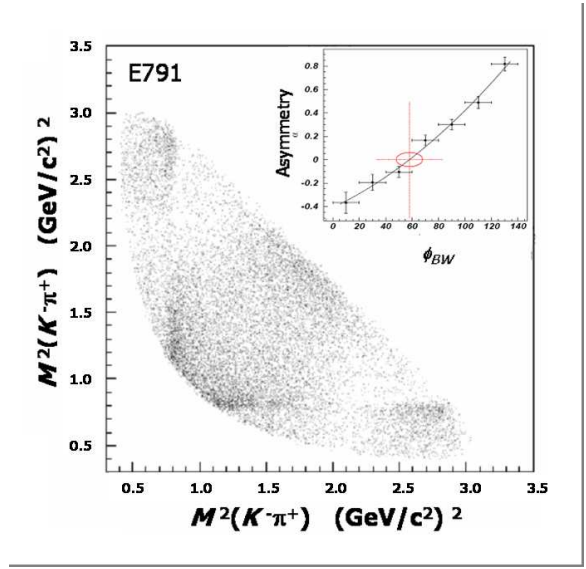,
          width=0.45\textwidth,
          angle=0}
\caption{Dalitz plot for $\Dp\to\Km\pip\pip$ decays.  Data are from E791
 \cite{Aitala:2002kr} and show distributions of squared invariant mass for
 one $\Km\pip$ combination plotted against the other (symmetrized).  The
 inset shows the asymmetry in the cosine of the $\Km\pip$ helicity angle
 (between $\Km$ and the other $\pip$ in the $\Km\pip$ rest frame) as a
 function of $\phi_{\sst BW}$, the $K^*(890)$ Breit-Wigner phase.
} \label{fig:E791dp}
\end{figure}

A study of $\sim 15,000$ such decays by the E791 collaboration 
\cite{Aitala:2002kr}
provides an illustration.  The E791 Dalitz plot is shown in
Fig.~\ref{fig:E791dp} where significant $S$-$P$ interference is
evident from the asymmetry of the $K^*(890)$ bands.  This is plotted
in Fig.~\ref{fig:E791dp} vs. the Breit-Wigner phase
$\phi_{BW}=\tan^{-1}M_0\Gamma(M_0)/M_0^2-s)$, where $M_0$ is the mass
and $\Gamma(\sqrt s)$ the invariant mass-dependent width of the
$K^*(890)$.  The asymmetry is zero when $\phi_{\sst BW}=56^{\circ}$,
almost $80^{\circ}$ below the phase found in the LASS data
\cite{Aston:1987ir}.
This shows that the phase of $Q_{K\pi}(s)$ at this invariant mass
is $\sim -80^{\circ}$ and is entered in Table~\ref{tab:wmd}.

\subsubsection{Isobar Model Fits}
In the earliest analyses of these decays, a model with interfering
resonances like the $K^*(890)$, the $L=0$ $K^*_0(1430)$ and a constant
non-resonant ``$NR$" 3-body amplitude could account for the voids and
asymmtries observed in the Dalitz plot.  With their larger sample,
the E791 isobar model analysis showed that additional
structure in the $S$-wave was required to achieve an acceptable fit,
and the addition of a $\kappa(800)$ Breit-Wigner isobar, which
interfered destructively with the $NR$ term, worked well.

The ``isobar model" description of the $L=0, 1$ and $2$ wave
amplitudes $F_L$ in the $\Km\pip$ systems for this fit can be
summarized as:
\begin{eqnarray}
  F_0(s) &=& c_{00}
             + \alpha_{10}BW_{K^*_0(1430)}(s) \nonumber \\
         & & + \alpha_{20}BW_{\kappa(800)}(s)
  \label{eq:s_isobar}
  \\
  F_1(s) &=& \alpha_{11}BW_{K^*_1(890)}(s) \nonumber \\
         & & + \alpha_{21}BW_{K^*_0(1688)}(s)
  \label{eq:p_isobar}
  \\
  %\label{eq:p_isobar}
  F_2(s) &=& \alpha_{21}BW_{K^*_2(140)}
  \label{eq:d_isobar}
\end{eqnarray}
where the $BW(s)$ are relativistic Breit-Wigner functions with
$s$-dependent widths, and the $\alpha_{iL}$ are complex coefficients
determined in the fit.  The overall phase was defined by setting
$\alpha_{11}=1.0$, and $\alpha_{00}$ was the $NR$ term.  

Two further isobar model analyses of this decay mode were recently
made, one by FOCUS
\cite{Pennington:2007se}
and the other by CLEOc
\cite{Bonvicini:2007nn}.
Each used samples $\sim 3.5$ times larger.  The conclusions, and
estimates of the resonant fractions
\footnote{The fraction for each resonance or $NR$ sub process is
defined in ref.~\cite{Aitala:2002kr}.}, of both were 
in good general agreement with E791.

Parameters for $\kappa$ and $K^*_0(1430)$ $S$-wave Breit Wigner isobars
are compared for the three experiments in Table~\ref{tab:kappapars}.
The $K^*_0(1430)$ parameters in this model disagree significantly with
those obtained by the LASS experiment or with the World average
\cite{pdg06}.
There is general agreement that this description of the $S$-wave,
described by Eq.~(\ref{eq:s_isobar}), with two broad, Breit Wigner
resonances, one of which is also near threshold, is theoretically
problematic and could account for this discrepancy.  It would be
virtually certain that this amplitude would have an $s$-dependent
phase that would differ from the Watson theorem expectation.
\begin{table}[hbt]
\begin{center}
\caption{Breit Wigner Parameters for the $\Km\pip$ $S$-wave Isobar
 States.  All quantities are in MeV/c$^2$.}
 \[
 \begin{array}{|l|l|r||r|r|r|}
 \hline
   %\mco{2}{c}{ }{\hss\hbox{\bf Parameters}\hss}
   \mco{2}{|c|}{ }
   & \hss\hbox{\bf E791}\hss
   & \hss\hbox{\bf Focus}\hss
   & \hss\hbox{\bf CLEO c}\hss
   & \hss\hbox{\bf PDG}\hss \\[3pt]
 \hline &&&&&\\[-10pt]
   \kappa          & M_{\circ}
                   &  797\pm 19\pm 42
                   &  883\pm 13 
                   &  805\pm 11
                   &  672\pm 40 \\
                   & \Gamma_{\circ}
                   &  410\pm  43\pm  85 
                   &  355\pm  13
                   &  453\pm 21
                   &  550\pm 34 \\[6pt]
 \hline &&&&&\\[-8pt]
   K^*_{\circ}     & M_{\circ}
                   & 1459\pm  7 \pm 12
                   & 1461\pm  4 
                   & 1461\pm  3
                   & 1414\pm ~6 \\
                   & \Gamma_{\circ}
                   &  175\pm 12\pm 12 
                   &  177\pm ~8
                   &  169\pm  5
                   &  290\pm 21 \\
 \hline
 \end{array}
 \]
\end{center}
\label{tab:kappapars}
\end{table}

\subsubsection{\boldmath Model-Independent Measurement}
A test of the Watson theorem requires a measurement of the
phase of $F(s)$ at several values of $s$ in a
model-independent way.  The first attempt to do this for the
$S$-wave for $\Km\pip$ produced in this decay mode was made
by the E791 collaboration
\cite{Aitala:2005yh}.

\begin{figure}[hbt]
\centering
\epsfig{file=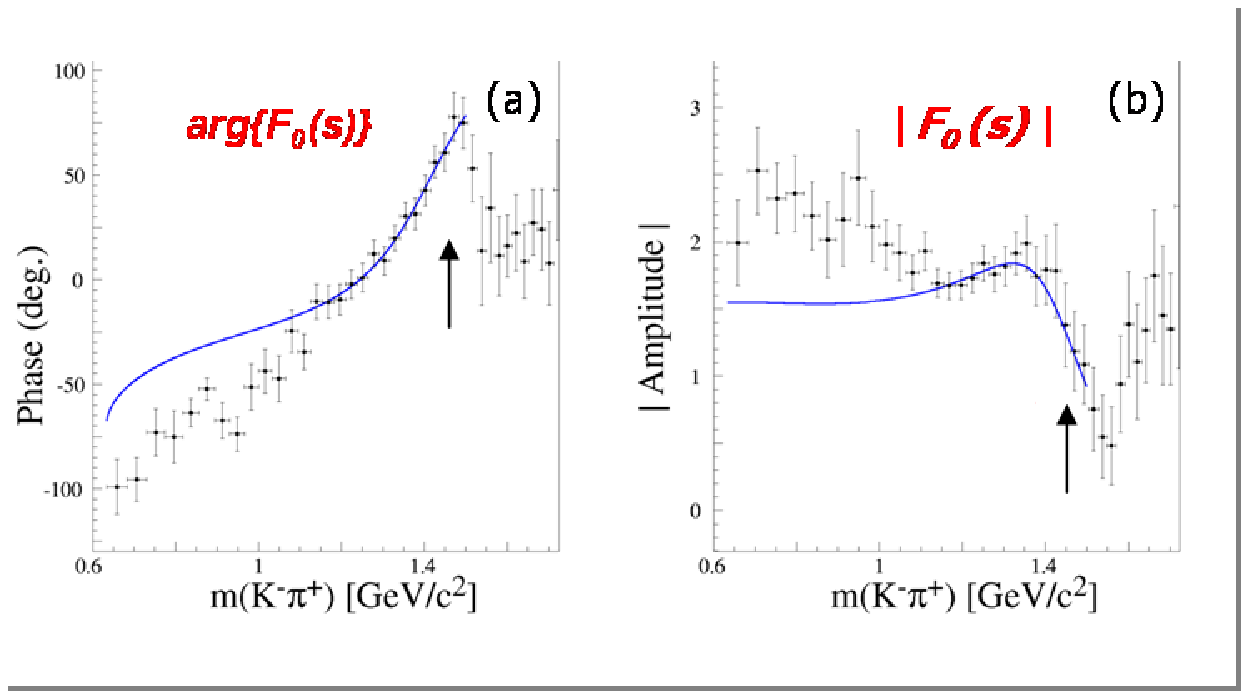,
          width=0.49\textwidth,
          angle=0}
\caption{(a) Phase and (b) magnitude of the decay amplitude $F_0(s)$
 at 38 discrete values of $s$ (squared invariant mass of the
 $\Km\pip$ systems from $\Dz\to\Km\pip\pip$ decays taken from
 Ref.~
 \cite{Aitala:2005yh}.  
 The LASS model in Eq.~(\ref{eq:lasswave}), shifted by $-75^{\circ}$ 
 and scaled to the region where $s>1.2$~(GeV/c$^2$)$^2$ is shown as 
 the blue, continuous curve.
} \label{fig:E791mipwa}
\end{figure}

They replaced the analytical function describing the $S$-wave
in Eq.~(\ref{eq:s_isobar}) by a set of 38 complex values at discrete
values for $s$, using a spline interpolation for other values.
The $P$- and $D$-waves were paramerized, as before, by the form
in Eqs.~(\ref{eq:p_isobar}) and (\ref{eq:d_isobar}) and the
coefficients $\alpha_{iL}$ were allowed to float.  A fit was
then made to determine the best values (magnitude and phase)
for each of the 38 $S$-wave points.  The result, shown in
Fig.~\ref{fig:E791mipwa}, is compared with the LASS model for the
$I=1/2$ $\Km\pip$ system in Eq.~(\ref{eq:lasswave}).  Agreement is
good in the $K^*_0(1430)$ region (above $\sim$1100~MeV/c$^2$)
after a shift in phase of $75^{\circ}$ and an arbitrary scale
factor are applied to the LASS amplitude.  A very significant
discrepancy is, however, seen for lower values of invariant
mass.

This might conform to our 4th expectation, but there are two
problems.  First, though $I=1/2$ $\Km\pip$ production probably
dominates, $I=3/2$ production in these decays cannot be excluded.
Second, the isobar model form in Eqs.~(\ref{eq:p_isobar}) and
(\ref{eq:d_isobar}) for the $P$- and $D$-waves, upon which this
result depends, is questionable. The $P$-wave contains more than
one Breit-Wigner, and both waves are assumed to be dominated by
resonant behaviour.  More importantly, neither wave is likely to
follow the Watson theorem, so a test of the $S$-wave alone cannot
be conclusive.

The CLEO collaboration
\cite{Bonvicini:2007nn}
has attempted to overcome the latter difficulty.  Using their
high purity sample of $\sim 60,000$ events, they proceeded in the
same way as E791, interpolating the $S$-wave between discrete
values of $s$, while parametrizing the $P$- and $D$-waves
as above.  Their results were very similar.  They then fixed the
$S$-wave and, using a similar procedure, fit the $P-$wave parametrized
in the same way.  Then they repeated this for the $D$-wave.  In
each step, only one wave was allowed to float with the others fixed.

This procedure can converge only by simultaneously floating all
waves at once, and this was not done.  Also, the phases have to be
defined in some part of the phase space (as little as possible) in
order for this to work.

It seems that a truly model-independent measurement of the $S$-wave
decay amplitudes in these hadron decays is desirable, but has yet
to be made, and will probably need to await the much larger data
samples of the $B$ factories, BES or the Panda experiment.

\subsubsection{\boldmath $I$-spin Analysis by the FOCUS Collaboration}
Using a sample of about $50,000$ events with 96\% purity, 
the FOCUS collaboration used the $K$-matrix method to separate
$I=1/2$ and $I=3/2$ production for this decay
\cite{Pennington:2007se}.

The $P$- and $D$-waves were described as in
Eqs.~(\ref{eq:p_isobar}) and (\ref{eq:d_isobar}), however, the $S$-wave
amplitude $F_0(s)$ in Eq.~(\ref{eq:s_isobar}) was replaced by the sum
of two terms, one for each $I$-spin.  Each was described by 
\begin{eqnarray}
  F_f(s)    &=& (I-i\rho K(s))^{-1}_{fk}P_k(s) \\
  \label{eq:FK}
  T_{kf}(s) &=& (I-i\rho K(s))^{-1}_{ki}K_{if}(s)   \\
  \label{eq:TK}
  Q_f(s)    &=& K^{-1}_{fk}(s)P_k(s).
  %\label{eq:TK}
\end{eqnarray}
Here, the amplitudes $F$, $T$, $P$ and $Q$ were introduced for
each $I$-spin and their indices $K$ and $f$ labelled intermediate
and final states ($1=K\pi$, 2=$\eta'K$).
% illustrated in Fig.~\ref{fig:Watson}
The production vector $P_k(s)$ describing the couplings of the $D$
decay to the two intermediate states was described by parameters
obtained from the fit, and $\rho_{kf}$ was the phase space matrix
for the two channels.

Real values for all elements of the $K$ matrices ($2\times 2$ for
$I=1/2$ and $1\times 1$ for $I=3/2$) were determined by a fit to
the LASS measurements of $T$.

The fit required a very large contribution from the $I=3/2$
$S$-wave ($40.50$\%) interfering destructively with $I=1/2$
($207.25$\%) giving a total $S$-wave ($83.23$\%).

\begin{figure*}[hbt]
\centering
\epsfig{file=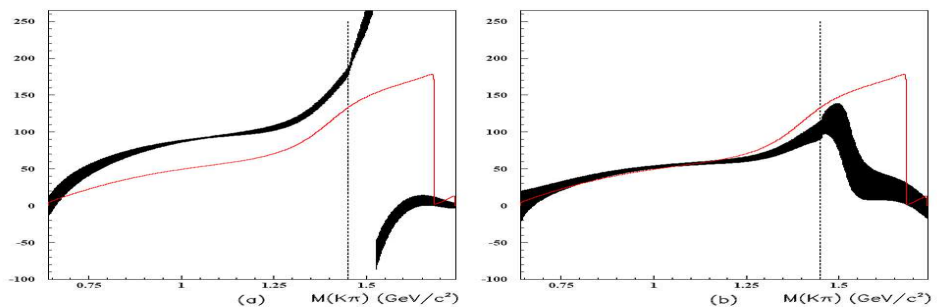,
          width=0.75\textwidth,
          height=0.25\textwidth,
          angle=0}
\caption{Decay amplitudes for $\Km\pip$ systems from $\Dp\to\Km\pip\pip$
 decays from 
 Ref.~\cite{Pennington:2007se}
 for a(a) the total $S$-wave and (b) the $I=1/2$ contribution.
 The black bands correspond to one standard deviation limits
 about the central fit in each of the amplitudes.
 Scattering amplitudes from a $K$-matrix fit to the LASS data
 \cite{Aston:1987ir}
 are shown as red continuous lines.
} \label{fig:FOCUSkpipi}
\end{figure*}

The phases of the resulting decay amplitudes, $F(s)$, are shown in
Fig.~\ref{fig:FOCUSkpipi}, with the $K$-matrix fit to the LASS
elastic scattering data for $I=1/2$ superimposed.  The phase of
the total amplitude $F(s)$, similar to that found by E791 and
by CLEOc, differs significantly from the LASS $I=1/2$ data.
The $I=1/2$ component is shown alone in 
Fig.~\ref{fig:FOCUSkpipi}(b) and shows better agreement.

The parametrization of the production vector was chosen
specifically to allow an $s$-dependence in its phase, thereby
allowing a deviation from the Watson theorem.  Some deviations
are, indeed, evident in Fig.~\ref{fig:FOCUSkpipi}(b), as the
invariant mass approaches the $K^*_0(1430)$ region.  It is
possible that this is due to the $K^*_0(1430)$ pole and the
onset of effects from the $K\eta'$ channel.  However, another
interpretation may be possible.  The phase shown in this figure
is shifted by an arbitrary amount to achieve good agreement at
the lower invariant masses.  Were a different shift in phase
applied, agreement would, in fact, be good at the high invariant
mass end and poor at low mass, consistent with our expectation
number 4.

This analysis attempted to make a valid comparison between
scattering and decay.  It appears premature, however, to
conclude that the Watson theorem works for these hadronic decays
especially because the same is not required for the $P$- nor
$D$-waves in this analysis.

\section{\boldmath $\pi\pi$ Production from Heavy Quark Meson Decays}
Decays of $\Dp\to\pim\pip\pip$ result in $S$-wave enriched
$\pip\pim$ systems.  Unfortunately, these decays are Cabibbo
suppressed.  Another decay rich in $S$-wave content is
$\Dz\to\Kz\pip\pim$.  This is partly Cabibbo favoured and partly
doubly suppressed, so is somewhat complex with many resonances
contributing.  Consequently, model-independent analyses of these 
decays have yet to be attempted.

\subsection{\boldmath $\Dp\to\pim\pip\pip$ Decays}
The largest sample yet studied comes from the CLEO experiment
\cite{Bonvicini:2007tc}.
It consists of only about $4,000$ events with a purity of
$\sim 95$\%.  It has been used to test a variety of models.

An isobar model fit to the Dalitz plot confirms, as does a 
similar analysis by FOCUS
\cite{Link:2003gb},
the E791 collaboration conclusion
\cite{Aitala:2000xu}
that structure in the low mass $\pip\pim$ system is well 
described by a scalar $\sigma(500)$ Breit Wigner destructively
interfering with a constant $NR$ amplitude.  The fit obtained
was of marginal quality.

CLEO tried several variations on the isobar model.  They included
an $I=2$ $\pip\pip$ $S$-wave contribution, slightly improving the
fit quality.  They also tried variations of $\pi\pi$ $S$-wave isobar
model.  In one, as suggested in 
Ref.~\cite{Oller:2004xm},
they replaced the $\sigma$ Breit-Wigner by a simple pole
\begin{eqnarray}
  1/(m_0^2-s-im_0\Gamma)~\to~1/(s_0-s)
\end{eqnarray}
where $s_0=(0.47-0.22 i)$.
Other $S$-wave models used were an amplitude based on the linear
sigma model
\cite{Schechter:2005we}
and an amplitude derived by N.Achamov described in the CLEO paper
Ref.~\cite{Bonvicini:2007tc}.
All models provided fits to the data slightly more acceptable than
the isobar model, but no clear distinction was obvious.

\subsection{\boldmath $\Dz\to\Kz\pip\pim$ Decays}
Isobar model fits to large samples have been made by \babar\
\cite{Aubert:2005iz}
and \belle\
\cite{Poluektov:2006ia}
collaborations.  They each require two $\sigma$ states in the $\pip\pi$
$S$-wave, one similar to that found in isobar fits to the previous
channel by E791, FOCUS and CLEOc, and the other at a mass of
$\sim 1.0$~GeV/c$^2$.  Both collaborations have also used
$K$-matrix parametrizations of the $S$-wave $\pip\pim$ amplitude,
obtaining slightly better fits.  These fits involve no assumptions
about $\sigma$ states, and they do enforce the Watson theorem in
this system.  However, the  other waves are those defined by the
isobar model, with no such constraint.  None of the fits so far
published have acceptable quality.

Further progress in this system may come from the larger data samples
form the $B$ factories, or from the next generation of charm factories.

\subsection{Other Channels}
A number of $B$ and $D$ decays in which $K\pi$ systems are
produced have now been published.  A comparison between the
general characteristics of the $S$-wave amplitudes observed and
those of the LASS scattering data are summarized in
Table~\ref{tab:wmd}
\cite{wmd:private}.
The two examples above are included and
represent the best examples of the validity of the Watson theorem.
Hopefully, an underlying pattern for these characteristics will
eventually emerge, provided that such decays are studied with this
goal in mind.  Any pattern is certainly not yet evident.

\begin{table}[hbt]
\begin{center}
\caption{Qualitative comparison between $K\pi$ $S$-wave decay
 amplitude $F_0(s)$ and the LASS Model in a variety of cases studied. 
 Three characteristics are compared.  The $S$-$P$ relative phase at
 the $K^*(890)$ mass.  The observed value minus that observed in
 LASS data, $\Delta\phi_{SP}$, is tabulated to the nearest
 15$^{\circ}$).  The general shape of $|F_0(s)|$ for $s$ values
 both below and above $s_0=1.0$~GeV/c$^2$ are also tabulated.}
\begin{tabular}{|ll|c|c|c|}
\hline  \mco{2}{|c|}{\textbf Decay Process} &
               \boldmath\bf $\phi_{SP}^{\circ}$
                &
               \boldmath $|F_0|$
                &
               \boldmath $|F_0|$
\\
                &
                &
               \textbf{approx}
                &
               \boldmath $s\!<\!s_0$
                &
               \boldmath $s\!>\!s_0$
\\
\hline $B^0\to J/\psi\Kp\pim$
 &\cite{Aubert:2004cp}~
 &
 $+180$ &
 Poorly &
 Similar
\\
 &&& defined & to LASS \\
\hline $B^+\to \Kp\pim\pip$
 &\cite{Aubert:2005ce}
 &
 $0$ &
 Unknown &
 Similar \\
 &&&         & to LASS \\
\hline $B^+\to \Kp\pim\rho$
 &\cite{Aubert:2006fs}
 &
 $+180$ &
 Unknown &
 Unknown \\
 &&&         &         \\
\hline $\Dz\to\Km\Kp\piz$
 &\cite{Mishra:2007dc}
 &
 $-90$ &
 Similar &
 Similar \\
 &&& to LASS & to LASS \\
\hline $\Dp\to\Km\pip\pip$
 &\cite{Aitala:2005yh}
 &
 $-75$ &
%\parbox{
 Very 
%\\ Steep rise \\towards threshold}
            &
 Similar \\
 &&& different & to LASS \\
\hline $D_s^+\to\Km\Kp\pip$
 &
 & $-90$ &
 Similar &
 Similar \\
 &&& to LASS & to LASS \\
\hline $\Dp\to\Km\pip\ell\nu$
 &\cite{Link:2005ge}
 &
 $0$ &
 Similar &
 Similar \\
 &&& to LASS & to LASS \\
\hline
\end{tabular}
\label{tab:wmd}
\end{center}
\end{table}

\section{Conclusions}
The most reliable data on $S$-wave amplitudes are still those
from LASS or CERN-Munich data on elastic scattering.  Data at
the lowest energies, however, are somewhat sparse.  It would
surely help in the understanding of the pole structure relevant
to the firm establishment of existence or otherwise of
light scalar $\kappa(800)$ or $\sigma(500)$ states to improve
on this situation.  It is difficult, at present, to see how
hadronic decays of $D$ or $B$ mesons will help since $I$-spin
considerations, and uncertainties in the $D$ form-factor, make
such measurements difficult.  It appears, therefore, that such
information is most likely to come from high statistics studies of
$D$ semi-leptonic decays, or decays of $B$ mesons to $J/\psi\Kp\pim$
(or $J/\psi\pip\pim$) from \babar\ or \belle\ or the new facilities
in BES or Darmstadt.

Progress in understanding ways to parametrize the $S$-wave decay
amplitudes in Dalitz plot analyses is slow, but the body of
information is growing.  The prize could be less systematic uncertainty
resulting from model dependence of such analyses in measurements of
important phenomena such as $\Dz$-$\Dzbar$ mixing or of the CKM
angle $\gamma$.

A positive trend is that more techniques beyond the isobar model,
with its known problems and strong model-dependence, are being
developed.

\begin{acknowledgments}
The author would like to thank the organizers for arranging this
workshop, and for the invitation to speak.  The support of the
National Science Foundation is also gratefully acknowledged.
\end{acknowledgments}

\bigskip % extra skip inserted

\bibliography{charm07_BrianMeadows}

%\begin{thebibliography}{99} % Use for 10-99 references
%
% \bibitem{fpcp07}  \phantom{http://www-f9.ijs.si/fpcp07/}
%\end{thebibliography}
%
\end{document}

%% file: babar.tex
\newcommand{\BaBar}{\mbox{$\mathrm{Ba\overline Bar}$}~}
\let\Babar=\BaBar
\let\mathrm=\rm
\newcommand{\PEP}{\mbox{$\mathrm{PEP}$}~}
\let\pep=\PEP
 
\newcommand{\hdr}[1]{{\Black{%
\small\phantom{.}\hfill#1\quad\today}}\newline\vspace{5mm}}
\newcommand{\sectn}[1]{\bigskip\noindent\underline{#1}\newline\noindent}

\newcommand{\CP}{CP\hskip-.5\em /}
\newcommand{\jpsi}{J/\psi}
\let\Jpsi=\jpsi
\newcommand{\KS}{K^{0}_s}
\newcommand{\KL}{K^{0}_L}
\newcommand{\Kz}{K^{0}}
\newcommand{\Kzbar}{\overline{K^{0}}}
\newcommand{\Kbar}{\overline{K}}
\newcommand{\Kst}{K^{\ast\circ}}
\let\Kstz=\Kst
\newcommand{\Kstbar}{\overline{K}^{\ast\circ}}
\let\Kstzbar=\Kstbar
\newcommand{\pz}{\pi^{0}}
\newcommand{\qbar}{\overline{q}}

\newcommand{\etal}{{\sl etal}~}
\newcommand{\wkph}{\delta_{\scriptscriptstyle W}}
\newcommand{\wkmix}{\delta_{\scriptscriptstyle M}}
\newcommand{\wkpp}{\wkph^{\prime}}
\newcommand{\wkpd}{\delta_{\scriptscriptstyle D}}
\newcommand{\wkpdp}{\wkpd^{\prime}}
\newcommand{\stph}{\delta_s}
\newcommand{\stpp}{\delta_s^{\prime}}
\newcommand{\Bz}{B^{0}}
\newcommand{\Bzbar}{\overline{B^{0}}}
\let\bz=\Bz
\let\bzbar=\Bzbar
\newcommand{\Bbar}{\overline{B}}
\newcommand{\Dz}{D^{0}}
\newcommand{\Dzbar}{\overline{D^{0}}}
\newcommand{\DMT}{{\Delta m\over 2}}
\newcommand{\fbar}{\overline{f}}

% **************************************************************************

\newcommand{\bott}{\vfill\phantom{.}}

\newcommand{\rhoz}{\rho^{0}}
\newcommand{\etp}{\eta^{\prime}}
\newcommand{\mz}{m_{\circ}}
\newcommand{\fz}{f_{0}}
\newcommand{\az}{a_{0}}
\newcommand{\ubar}{\overline{u}}
\newcommand{\dbar}{\overline{d}}
\newcommand{\sbar}{\overline{s}}
\newcommand{\bbar}{\overline{b}}
\newcommand{\cbar}{\overline{c}}
\newcommand{\tbar}{\overline{t}}
\newcommand{\piz}{\pi^{0}}
\newcommand{\pip}{\pi^+}
\newcommand{\pim}{\pi^-}
\newcommand{\Kp}{K^+}
\newcommand{\Km}{K^-}
\newcommand{\Dsp}{D_s^+}
\newcommand{\Dp}{D^+}
\newcommand{\Dst}{D^{\ast}}
\newcommand{\Dpst}{D^{\ast+}}
\newcommand{\Xcp}{\Xi_c^+}
\newcommand{\Xcz}{\Xi_c^{0}}

\newcommand{\Aj}{{\cal A_{\rm j}}}
\newcommand{\llik}{{\cal L}}
\newcommand{\aj}{\Red{a_{\rm j}}}
\newcommand{\deltj}{\Red{\delta_j}}
\newcommand{\fj}{\Red{f_{\rm j}}}
\newcommand{\bi}{\Red{b_{\rm i}}}

%% file: mytex.tex
\newcommand{\st}{\scriptstyle}
\newcommand{\sst}{\scriptscriptstyle}
\newcommand{\ie}{{\sl i.e.~}}
\newcommand{\eg}{{\sl e.g.~}}
\newcommand{\etc}{{\sl etc~}}
\newcommand{\vs}{{\sl vs.~}}
\newcommand{\half}{\ensuremath{{1\over 2}}}
\newcommand{\z}{\ensuremath{_{\circ}}}
\newcommand{\amu}{~{\rm u}}
\newcommand{\mps}{~{\rm m/s}}
\newcommand{\Hz}{~{\rm Hz}}
\newcommand{\eV}{~\hbox{eV}\ }
\newcommand{\eVcc}{~\ensuremath{\hbox{eV}/c^2}}
\newcommand{\kg}{~{\rm kg}}
\newcommand{\meter}{~{\rm m}}
\newcommand{\J}{~{\rm J}}
\newcommand{\K}{~{\rm K}}
\newcommand{\epsz}{\ensuremath{\epsilon_{\circ}}}
\newcommand{\muz}{\ensuremath{\mu_{\circ}}}
\newcommand{\grad}{\ensuremath{\vec\nabla}}
\newcommand{\curl}{\ensuremath{\vec\nabla\times\vec}}
\newcommand{\boombox}[2]{\centerline{%
            \shadowbox{\parbox{#1}{#2}}}}
\newcommand{\sidetxtfig}[2]{\leftline{%
\mbox{%
\begin{minipage}[b]{3.0in}{#2}%
\end{minipage}
\epsfig{file=#1,width=4.0in}
} % end of \mbox{
} % end of \leftline{%
}%
\newcommand{\sidefigtxt}[2]{\leftline{%
\parbox{0.49\hsize}{#1}%
\hskip4mm
\parbox{0.49\hsize}{#2}%
} % end of \leftline{%
} % end of \newcommand}{\sidefigtxt